\documentclass[prl,superscriptaddress,showpacs,twocolumn]{revtex4}
\usepackage{graphicx}
\usepackage{dcolumn}% Align table columns on decimal point
\usepackage{bm}% bold math
\usepackage{amsmath}
\usepackage{amssymb}
\usepackage{color}
\usepackage{comment}
\usepackage{ulem}
\usepackage{float}
\usepackage{verbatim}

\newcommand{\ket}[1]{\left|#1\right>}

\DeclareMathOperator{\arcsinh}{arcsinh}

\def \bk{{\bf k}}
\def \br{{\bf r}}
\def \bq{{\bf q}}
\def \bQ{{\bf Q}}

\def\Vkk{\frac{1}{|\bk-\bk'|^2}}

\def\T{T}

\def\R{\mathbb R}

\def\F{\mathrm{{\mathcal F}}}

\def \FG{{\rm FG}}
\def \SDW{{\textsc{\tiny SDW}}}

\def \aK{{a_K^{}}}
\def \aV{{a_V^{}}}

\def \Fu{{\mathcal F}_\uparrow}
\def \Fd{{\mathcal F}_\downarrow}
\def\de{\delta\!E^\F_\SDW}
\def\epsfac{\exp\left(-\frac{\pi}{\sqrt 2}\sqrt {\gamma}\right)}

\begin{document}
\title{Upper bounds of spin-density wave energies in the homogeneous electron gas}
\begin{abstract}
Studying the jellium model in the Hartree-Fock approximation, Overhauser has shown that spin density waves (SDW) can lower the energy of the Fermi gas, but it is still unknown if these SDW are actually relevant for the phase diagram.
In this paper, we give a more complete description of SDW states. We show that a modification of the Overhauser ansatz explains the behavior of the jellium at high density compatible with previous Hartree-Fock simulations.
\end{abstract}
\author{F. Delyon}
\affiliation{CPHT, UMR 7644 of CNRS, \'Ecole Polytechnique, F-91128 Palaiseau Cedex, France}
\affiliation{LPTMC, UMR 7600 of CNRS, UPMC, Paris-Sorbonne, F-75252 Paris Cedex 05, France}
\author{B. Bernu}
\affiliation{LPTMC, UMR 7600 of CNRS, UPMC, Paris-Sorbonne, F-75252 Paris Cedex 05, France}
\author{L. Baguet}
\affiliation{LPTMC, UMR 7600 of CNRS, UPMC, Paris-Sorbonne, F-75252 Paris Cedex 05, France}
\author{M. Holzmann}
\affiliation{LPTMC, UMR 7600 of CNRS, UPMC, Paris-Sorbonne, F-75252 Paris Cedex 05, France}
\affiliation{LPMMC, UMR 5493 of CNRS, Universit\'e J. Fourier,  BP 166, F-38042 Grenoble Cedex, France}

\maketitle
 The simplest model of electronic structure is jellium -- electrons embedded in a homogeneous background of opposite charge such that the system is neutral.
This model is a good starting point to describe properties of simple metals such as sodium\cite{Na}.
At zero temperature, the only parameter of this model is the density $n$, or the dimensionless parameter $r_s=3/(4\pi n a_B^3)^{1/3}$, where $a_B=\hbar^2/(me^2)$ is the Bohr radius.
Within the Hartree-Fock approximation (HF), Overhauser has shown that the Fermi gas is unstable under a spin density wave (SDW)\cite{Overhauser,Vignale}.
Only recently, almost 50 years after Overhauser's prediction, explicit numerical estimates of the HF ground state have shown SDW evidences of
the electron gas in three\cite{Zhang-Ceperley,Kurth,HF3DEG,BaguetThesis} and two dimensions\cite{HF-2D,BaguetThesis}. Still, a quantitative estimate of the variation of the SDW amplitude and energy in the high density region $r_s \lesssim 1$ is missing\cite{refBernu2008}. In this letter, we generalize Overhauser's ansatz and provide a quantitative solution of this long-standing problem.

\begin{figure}
\begin{center}
\includegraphics[width=0.28\textwidth]{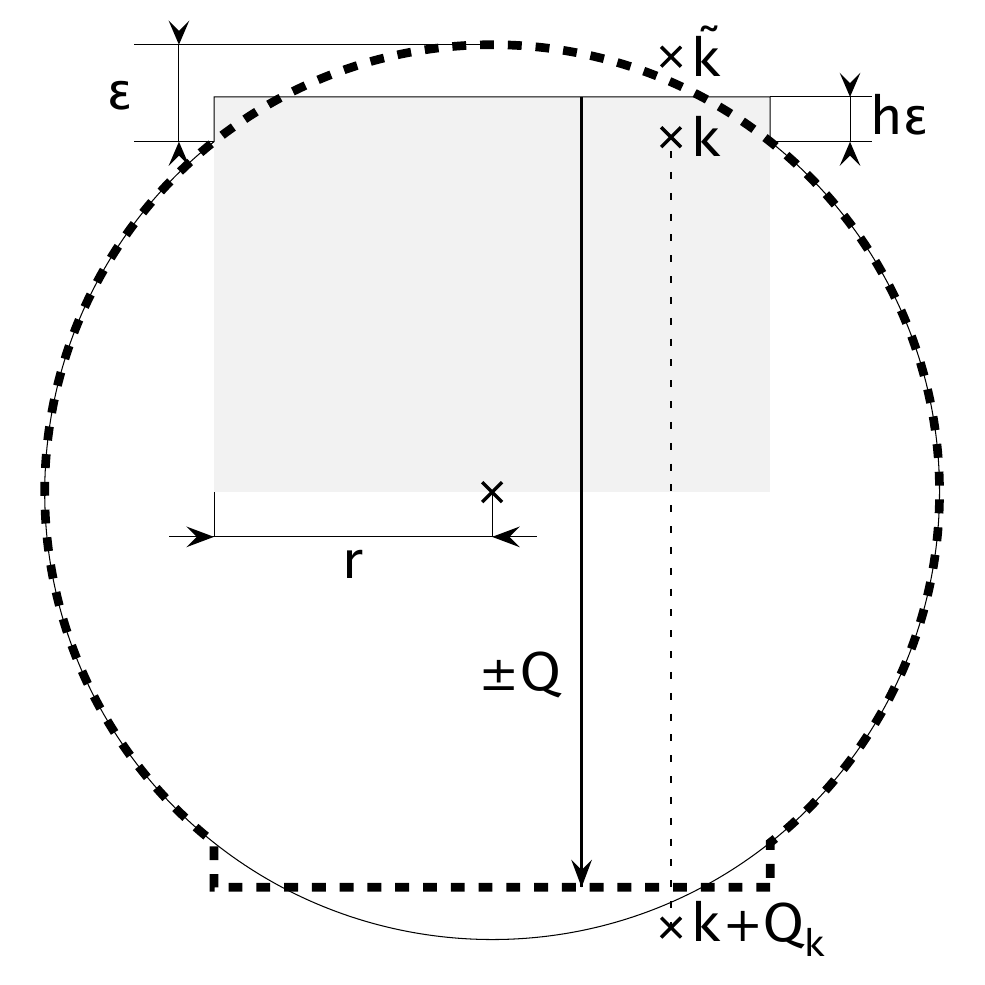}
\caption{Shape of occupied zones for spiral SDW. $\Fu$ (plain line) 
and $\Fd$ (dashed line) are built from truncated sphere plus a small cylinder.
%is the part of the sphere of radius 1 with the region $k_z<1-\epsilon+h$ (resp. $k_z>-1-\epsilon+h$).
A perturbed spin-up state is a superposition of a spin-up plane wave $e^{i\bk\br}$ with a spin-down plane wave $e^{i(\bk+\bQ)\br}$ with $Q=2(1-\epsilon+h\epsilon)$.
\label{fig33}
}
\end{center}
\end{figure}

The key point is to search for a solution in a non-perturbative way.
Indeed, small domains exist around the Fermi surface where the one-body states differ radically from a single plane wave.
These states of wave vector $\bk$ are coupled with the wave vector $\bk+\bQ_\bk$ where $\bQ_\bk$ is constant over each domain.
The larger is this domain, the larger will be the energy gain of the SDWs.
One way to enlarge this domain is to cut the top of sphere as explained by Overhauser\cite{Overhauser}.
In the following we show that adding a small cylinder on the top of the truncated sphere as shown in Fig.\ref{fig33} can increase the energy gain of the SDW by orders of magnitude compared to  Overhauser's ansatz.
Furthermore, we provide an explicit estimate of the energy gain of the SDW state. 
As we will see, the optimal size of these domains dramatically shrinks with increasing density, resulting in a extremely rapid decrease of the tiny SDW energy gain and
explaining the difficulties of observing SDW in the high density region.

Our semi-analytical results presented here are compared to recent HF results\cite{HF3DEG} obtained with periodic models.
Indeed, the Overhauser's ansatz is in fact
a periodic model (a crystal where the one-body states are limited to the first mode)  as soon as the set vectors $\bQ_\bk$ belong to a discrete lattice.
As the density increases the number of vectors $\bQ_\bk$ (and of domains around the Fermi sphere)
may also increases\cite{Overhauser} leading to a quasi-crystal which cannot be describe by a periodic model.

Let us mention that in this paper we focus on the SDW states. These states are easier to compute leading to simpler formulas since the density of charge is constant. Equivalent results may be obtained\cite{Overhauser,Vignale} for the charge density waves (CDW). 
%In particular, we show that the size effects are huge, explaining why such states are very difficult to exhibit at large density.

In the following, we outline the main steps in the calculation of the SDW energies. First we introduce the deformation of the Fermi surface generalizing Overhauser's model and describe the SDW ansatz for the single particle states. We then show how the optimal solution can be found by calculating the fixed point solution of a non-linear functional equation. The explicit results are then obtained by restricting to a one dimensional function and compared to the outcome of previous numerical simulations.

{\bf Fermi gas energy of the truncated sphere.}
Let us call $E_\FG$ the HF-energy of the Fermi gas where only plane wave states of wave vectors $\bk$ inside the Fermi sphere of radius $k_F$ are occupied.
Following Overhauser, in a first step  the Fermi sphere is deformed into a volume $\F$ as shown in Fig.\ref{fig33}, and its energy increase is denoted $\Delta E^\F_\FG=E_\FG^\F-E_\FG$.
Here, the subscript $\FG$ is used to point out that the many-body state is a Slater determinant of plane wave states inside the corresponding Fermi surface.
Using $k_F$ as unit of wave vectors, the sphere in 
Fig.\ref{fig33} has unit radius, and the deformation is characterised by a small parameter $\epsilon$ approaching zero as $r_s$ decreases. In order to keep the electron density constant, the deformed surface in the figure
must be scaled  by $R$ \cite{Rh} such that  $R^3\int_{\F_\uparrow}d\bk=R^3\int_{\F_\downarrow}d\bk=4\pi/3$.
The Fermi gas energy per particle in Hartree units ($Ha=\hbar^2/(ma_B^2)$) is
\begin{align}
\label{efg}
	E_\FG^\F&=\frac{a_K R^5}{ r_s^2}K_\FG
		- \frac {a_V R^4} {r_s}V_\FG\\
\label{efgK}
	K_\FG&=\int_{\F_\uparrow+\F_\downarrow}d\bk\ k^2\\
\label{efgV}
	V_\FG&=\int_{\F_\uparrow\times \F_\uparrow +\F_\downarrow\times\F_\downarrow}d\bk\,d\bk'\ \Vkk
%&=\frac{2}{ r_s^2}(\frac 45- {r_s})
\end{align}
with 
 $a_V=\frac{3}{32\pi^3}\left(\frac{9\pi}4\right)^{1/3}$, $a_K/a_V= 2\pi^2\left(\frac{9\pi}4\right)^{1/3}\approx 37.9$. 
 % for polarized: a_K/a_V --> a_K/a_V * \sqrt[3]2
% $a_V=3\alpha/(32\pi^3)$, $a_K=a_V 2\pi^2\alpha\approx 37.9\, a_V$ ($\alpha=(9\pi/4)^{1/3}$)
% and $\bk$ is in $k_F$ units

\begin{figure}
\begin{center}
\includegraphics[width=0.4\textwidth]{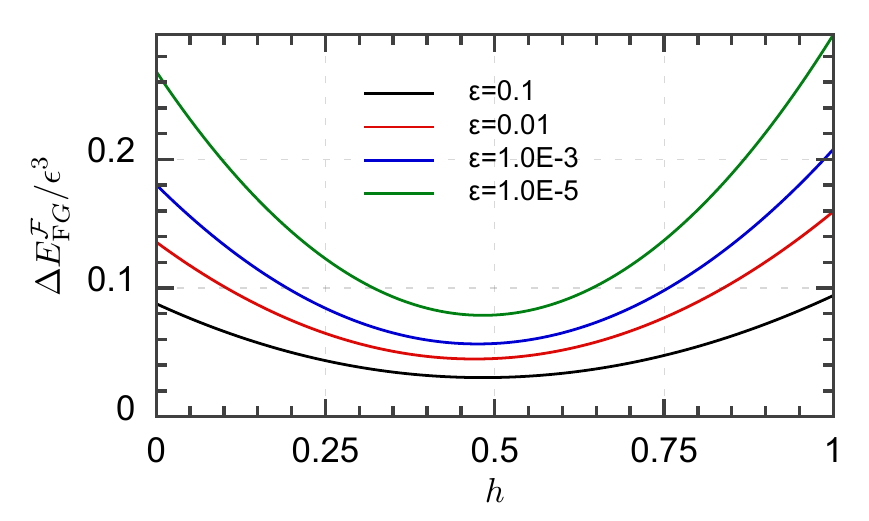}
\caption{$\Delta E^\F_\FG$ as a function of $h$ for different values of $\epsilon$ at $r_s=4$.
\label{EFG}
}
\end{center}
\end{figure}
The energy change, $\Delta E^\F_\FG$, can be computed by direct integration (see Fig.\ref{EFG}) and gives at the leading order in $\epsilon$\cite{supp}:
\begin{align}
\label{enFG}
	\Delta E^\F_\FG\approx& \frac{2\pi^2\aV\epsilon^3}{r_s}\left[ \alpha (\gamma-1)-\frac {1}{9}
	+h\right]
\end{align}
with $\alpha=2  (h-\frac 12)^{2}+\frac 16$ and $\gamma=\ln\frac2\epsilon+\frac{\aK}{\aV \pi r_s}$. For small $r_s$, $\epsilon$ is small and thus $\gamma$ is large.
The minimum of $\Delta E^\F_\FG$ with respect to $h$ is for
$h=\frac12-\frac1{4(\gamma-1)}\approx \frac12$,
%\begin{align}
%	h=\epsilon(\frac12-\frac1{4\gamma})%=\frac12-\frac14\frac1{\ln\frac2\epsilon+\frac{12.06}{r_s}-1}
%\end{align}
%and we have
%\begin{align}
%\label{enerfg}
%	\Delta E^\F_\FG\approx& \frac{\pi^2\aV\epsilon^3}{r_s}\left(\frac\gamma3+\frac 7{9}-\frac 1{4\gamma}\right) 
%\end{align}
and, for small $r_s$, $\Delta E^\F_\FG$ is four times smaller than in the Overhauser case ($h=0$).
 
{\bf The spin density waves.}
In a second step, the SDW are obtained by replacing a plane wave $\ket{\bk,\uparrow}$ by $a_\bk\ket{\bk,\uparrow}+b_\bk\ket{\bk+\bQ_\bk,\downarrow}$  ($|a_\bk|^2+|b_\bk|^2=1$), for $\bk$ in $\Fu$. Symmetrically, for $\bk$ in $\Fd$, a plane wave $\ket{\bk,\downarrow}$ is replaced by the combination $a'_\bk \ket{\bk,\downarrow}+b'_\bk\ket{\bk+\bQ_\bk,\uparrow}$ with $a'_\bk=a_{-\bk}$ and $b'_\bk=b_{-\bk}$.
We choose $a_\bk$ real and positive and in the following we assume that $b_\bk$'s are also positive and that $b_\bk<a_\bk$\cite{bpositive}.

%{\fd The resulting state is a periodic crystal with a modulation  given by $\bQ_\bk$. In the following, we compare with HF crystals with the same modulations in modulus taking into account 
%the number of modulations in the HF case.}

As in Fig.\ref{fig33}, $\bQ_\bk$ is such that for $\bk$ in $\Fu$, $\bk+\bQ_\bk$ does not belong to $\Fd$, and for $\bk$ in $\Fd$, $\bk+\bQ_\bk$ does not belong to $\Fu$.
The  energy change is given by:
\begin{align}
\label{eSDW}
	\Delta E^\F_\SDW=&\frac{2a_K R^5}{ r_s^2}K_\SDW
		- \frac {2a_V R^4} {r_s}V_\SDW\\
	K_\SDW=&\int_{\F_\uparrow}d\bk\ (\|\bk+\bQ_\bk\|^2-k^2)b_\bk^2\\
\nonumber
	V_\SDW=&-\int_{\Fu\times \Fu}\!\!\!\!d\bk d\bk'\ \frac {(a_\bk b_{\bk'}-b_\bk a_{\bk'})^2}{\|\bk-\bk'\|^2}\\
	%\vc_{\bk- \bk'}\\
\label{potSDW}
&+\int_{\Fu\times \Fu}\!\!\!\!d\bk d\bk'\ \frac {(a_\bk b_{\bk'}+b_\bk a_{\bk'})^2}{\|\widetilde \bk-\bk'\|^2}
%\vc_{ \tilde\bk- \bk'}
%&=\frac{2}{ r_s^2}(\frac 45- {r_s})
\end{align}
with $\widetilde k_z=Q-k_z$ (see Fig.\ref{fig33}). 
%We define 
Using the linear symmetric operators $\T^\pm$:
\begin{align}
\label{defT}
%	(\T^\pm f)(\bk)&=\int_{\F_\uparrow}d\bk'\ (\vc_{\bk-\bk'}\pm \vc_{\widetilde \bk-\bk'})f(\bk'),
	(\T^\pm f)(\bk)&=\int_{\F_\uparrow}\!\!\!\!d\bk'\!\left(\frac 1{\|\bk-\bk'\|^2}\pm \frac 1{\|\widetilde \bk-\bk'\|^2}\right)f(\bk'),
	%(T_+f)(\bk)&=\int_{\F_\uparrow}d\bk'\ (\vc_{\bk-\bk'}+\vc_{\tilde \bk-\bk'})f(\bk').
\end{align}
Eq.\ref{eSDW} rewrites:
\begin{align}
%\nonumber
\label{potSDW2}
	%\Delta E^\F_\SDW&= 2(\Vm,b^2)-(\T^-b^2,b^2)-(\T^+ab,ab)
	%V_\SDW=-2(a^2,\T^-b^2)-2(\T^+ab,ab)
	\Delta E^\F_\SDW&\!=\!\frac {4a_V R^4} {r_s}\left(2(\kappa,b^2)\!+\!(\T^-a^2,b^2)\!-\!(\T^+ab,ab)\right)
\end{align}
where  $(f,g)$ is the scalar product $\int_{\F_\uparrow} \!f g$, %$A_V= \frac {4a_V R^4} {r_s}$, 
and $2\kappa(k)=\frac{a_KR}{2 a_Vr_s}Q(Q-2k_z)\ge 0$. In Eq.\ref{potSDW2} the difference between $1$ and $R$ is negligible and in the following we set $R=1$.\
 %$\xi(\bk)\equiv \xi_\bk=a_\bk b_\bk=b_\bk\sqrt{1-b_\bk^2}$,
 
 {\bf Optimal solution.}
 From the variations of Eq.\ref{potSDW2} with respect to $b_\bk$,  the optimal function $b$ satisfies:
\begin{align}
\label{de}
	2b \kappa+b\T^-(a^2-b^2)=\frac{a^2-b^2}{a}\T^+ab
\end{align}
Setting $\xi=ab$ and  using $b^2\leq\frac12$, $a^2-b^2=\sqrt{1-4\xi^2}$, Eq.\ref{de} rewrites  $\xi=J(\xi)$ where:
\begin{align}
\label{eqxi}
	J(\xi)=\frac 12\frac {\T^+\xi}{\sqrt{\left(\kappa+\T^-\sqrt{1/4-\xi^2}\right)^2+(\T^+\xi)^2}}.
\end{align}
%From Eq.\ref{de}, the optimal energy is given by:
%\begin{align}
%\label{desimple}
%	\Delta E^\F_\SDW&= A_V\left((b^2,\T^-b^2)-(\T^+\xi,\frac{b^3}{a})\right).
%\end{align}

Thus the point is now to find the fixed points of the operator $J$.
The Fermi gas ($\xi=0$) is a trivial fixed point.
By definition $0\le \xi\le \frac 12$, and from Eq.\ref{eqxi}, we see that $0\le J(\xi)\le \frac 12$.
%Moreover, we will see that at the border of the cylinder ($k_z=Q/2$) we have $\xi=a^2=b^2=1/2$.
We claim that starting with $\xi=1/2$ and iterating the process $\xi\to J(\xi)$ leads to a non-trivial fixed point satisfying:
\begin{align}
\xi(k_z=Q/2)=\frac 12.
\end{align}
Indeed, by (\ref{defT}) the kernels of $\T^\pm$ are positive, thus  $\T^\pm$ are positivity preserving linear operators:
if $\xi\ge \xi'$, then $\T^+ \xi\ge \T^+ \xi'$; similarly  $\T^-\sqrt{1/4-\xi^2}<\T^-\sqrt{1/4-\xi'^2}$,
and consequently:
\begin{align}
\label{propk}
\xi\ge \xi' \Longrightarrow J(\xi)\ge J(\xi').
\end{align}
Thus starting with $\xi_0=1/2$, we have $J(\xi_0)\le \xi_0$ and setting $\xi_n=J(\xi_{n-1})$, $\xi_n$ is a decreasing sequence of positive functions and thus converges to a fixed point $\xi_\infty$. 
\begin{comment}
On the other hand, let $C_\delta$ be the cylinder $C=\{\bk: k_x^2+k_y^2<r^2=2\epsilon, k_z>Q/2-\delta\}$ and let define $\zeta_0$ as
\begin{align}
\zeta_0=\frac \eta 2\frac 1{\sqrt{1+\alpha^2(k_z-Q/2)^2}}\ \ \eta<1
\end{align}
for $\bk \in C_\delta$ and $\xi_0=0$ otherwise.

For  $k_z,k'_z\in C_\delta$, the kernel $\T^+(k,k')$ is larger than $\frac {C_1}{\epsilon}$ as soon as  $\delta$ is at most of order $\sqrt\epsilon$; thus for $\bk\in C_\delta$
\begin{align}
(\T^+\zeta_0)(\bk)\ge \frac {\eta C_1}{2{\epsilon}}\pi r^2\frac 1\alpha \log(\delta\alpha)
= \pi  \eta C_1\frac {1}\alpha \log(\delta\alpha)
\end{align}
provided that $\delta\alpha>1$
and 
\begin{align}
J\zeta_0\ge\frac 12\frac 1{\sqrt{1+(\frac \alpha{2\pi \eta C_1\log(\delta\alpha)})^2(\kappa+T^-\sqrt{1-4\xi^2})^2}}
\end{align}
Now $\kappa$ is a linear function of $(k_z-Q/2)$ and $T^-\sqrt{1-4\xi^2}=0$ at $k_z=Q/2$. The technical point is to show that the derivative of $T^-\sqrt{1-4\xi^2}$ is bounded.
In this case:
\begin{align}
J\zeta_0\ge\frac 12\frac 1{\sqrt{1+(\frac {\alpha C_2}{2\pi \log(\delta\alpha)})^2(k_z-Q/2})^2}
\end{align}
Thus $J\zeta_0\ge \zeta_0$ as soon as 
\begin{align}
\frac { C_2}{2\pi }\le\log(\delta\alpha)
\end{align}
That is, we can choose $\delta=\sqrt\epsilon$ and $\alpha$ of order $1/\sqrt\epsilon$. As above we can build the increasing sequence $\zeta_n=J(\zeta_{n-1})$
and thus $\zeta_0\le \zeta_\infty\le\xi_\infty$ and the fixed point  $\xi_\infty$ is non zero and satisfies $\xi_\infty(k_z=Q/2)=1/2$.
\end{comment}

{\bf 1-D approximation.} Now we impose that  $b_k$ (thus $\xi_\bk$) is non zero only in the cylinder ${\mathcal C}$ corresponding to the gray region of Fig.\ref{fig33} where it depends only on $k_z$: ${\mathcal C}=\{\bk: k_x^2+k_y^2\le r^2=1-(1-\epsilon)^2\approx 2\epsilon, 0\le k_z\le Q\}$.
%The SDW energy does not depend on $h$ (at the leading order) and here we choose $R=1$ i.e $h\approx 1/2$.
As we shall see below, $b_\bk$ differs from zero only in the neighborhood of the top disk of $\F_\uparrow$ (and its symmetric for $\F_\downarrow$).
%as $r_s$ goes to zero this restriction becomes in force since $\epsilon$ goes to  zero and $b_\bk$ differs from zero only in the neighborhood of the two 
%disks of the truncated spheres.
In any case, these restrictions always provides an upper bound for the energy of the SDW.

First, for the second term of Eq.\ref{potSDW2}, we have:
\begin{align}
(a^2,\T^-b^2)=(b^2,\T^-a^2)=(b^2,\T^-1)-(b^2,\T^-b^2)
\end{align}
From Eq.\ref{defT}, $\T^-1=v_\F(\bk)-v_\F(\widetilde \bk)$ where $v_\F$ is the potential induced by the truncated sphere.
In the spherical case, the potential of the unit sphere is given by:
$$v(\bk)=2\pi +\pi\frac{1-k^2}k\ln\frac{1+k}{|1-k|}$$
In this case, for $k$ close to 1 ($\bk$ and $\widetilde \bk$ are close and near the unit sphere and $\widetilde k_z=2-k_z$), $v(\bk)-v(\widetilde \bk)\approx -4\pi (1-k) \log(\frac{1-k}2)$ .
This singular behavior is associated to the discontinuity of the density (in $\bk$-space).
An analytic solution is provided for the truncated sphere \cite{supp}.
This solution has the same behavior except that $1-k$ has to be replaced 
by the distance of $\bk$ to the discontinuity of the density, i.e. the top disk of $\F_\uparrow$:
\begin{align}
\label{T1}
(\T^-1)(\bk)\approx-4\pi (Q/2-k_z)\log\left(\frac {|Q/2-k_z|}2\right)
\end{align}
provided that $|Q/2-k_z|\ll 1$.
For $h>0$, Eq.\ref{T1} is still valid\cite{supp} except in a small neighborhood of the edge of the top disk.
In the following we neglect this effect and apply Eq.\ref{T1} also for $h>0$.

Using the scaled distance $x=(Q/2-k_z)/r$, 
 \begin{align}
2\kappa+\T^-1=2\pi r(\gamma x-2x\log(x)),
\end{align}
and integrating over $\bq=(k_x,k_y)$, Eq.\ref{potSDW2} becomes:
\begin{align}
\label{Ex0}
	\Delta E^\F_\SDW&= \frac{4\pi a_Vr^4}{r_s} \,\de\\
\label{Ex}
	\de&=2\pi(\gamma x-2 x\log(x),b^2)-(\T^-b^2,b^2)-(\T^+\xi,\xi)
%	\de&=2(\Wm,b^2)-(w_-b^2,b^2)-(w_+\xi,\xi)
%\de&=2\int dx\,W_-(x)b(x)^2\\
%&-\int \,  dx (w_-b^2)(x)b(x)^2-\int \,  dx(w_+\xi)(x)\xi(x)
%&=\frac{2}{ r_s^2}(\frac 45- {r_s})
\end{align}
where the scalar product is now $(f,g)=\int _{x>0}\!dx\,f(x) g(x)$, 
and $\T^\pm$ become in terms of $x$:
\begin{align}
	(\T^\pm f)(x)=&\pi\!\! \int_0^{1/r} \!\!\!\!  dx'\left(G(x-x')\pm G(x+x')\right)f(x')\\
\label{Gxdef}
	G(x)=&\frac 1{\pi^2 r^2}\int_{q^2,{q'}^2<r^2}\!\!\!\!\!\!\!\!\!d\bq\,d\bq'\frac{1}{r^2x^2+(\bq-\bq')^2}\\
	%&=\frac 1{\pi^2 }\int_{q^2,{q'}^2<1}d\bq\,d\bq'\frac{1}{x^2+(\bq-\bq')^2}\\
\label{Gx}
	=&2\ln\left[1\!+\!\frac 2{|x|u}\right]\!-\!\frac {4}{u^2}, \  u=	|x|\!+\!\sqrt{x^2\!+\!4}
%	&=-\frac {4}{(|x|+\sqrt{x^2+4})^2}\\
%	&+2\ln(1+\frac 2{|x|(|x|+\sqrt{x^2+4})})
\end{align}

In fact, for small $r_s$, the term $(\T^-b^2,b^2)$ may be neglected in Eq.\ref{Ex} \cite{supp}. In any case, since $\T^-$ is a positive operator, we get an upper bound for the energy and the variation of the resulting upper bound leads to $\xi=J\xi$ where $J$ is now an operator on the positive functions on $\R^+$:
\begin{align}
\label{eqxi2}
	J(\xi)=\frac 12\frac {\T^+\xi}{\sqrt{\pi^2\left(\gamma x-2 x\log(x)\right)^2+(\T^+\xi)^2}}.
\end{align}
%Similar as before, this fixed point equation can also be solved by iteration.
As above, the fixed point of Eq.\ref{eqxi2} can be easily found by iteration.

\begin{figure}
\begin{center}
\includegraphics[width=0.4\textwidth]{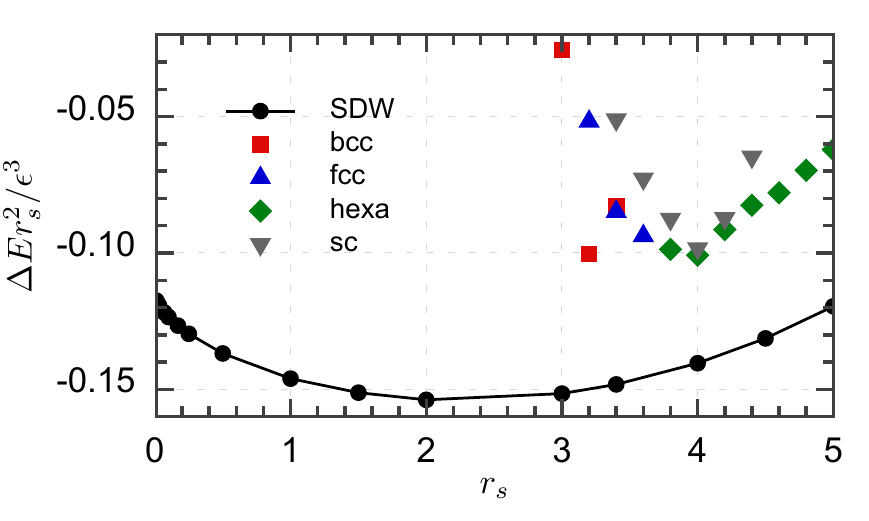}
\caption{Renormalized energy  $\Delta Er_s^2/\epsilon^3$ as a function of the density.  The dashed-dotted line stands for the analytical solution: $\Delta Er_s^2/\epsilon^3=-0.115$. 
$\blacktriangle$  stand for the SDW simulations. 
Others symbols stand HF energies (see Fig.5 of \cite{HF3DEG}).
\label{E}
}
\end{center}
\end{figure}

Thereafter, for fixed $r_s$ the total energy variation $\Delta E(\epsilon,h)=\Delta E^\F_\FG(\epsilon,h)+\Delta E^\F_\SDW(\epsilon)$ is computed and optimized with respect to $\epsilon$.
For $r_s=3$ about 20 iterations of the operator $J$ are required and about 100 iterations for $r_s=0.01$ (see $\blacktriangle$ symbols in Fig.\ref{E}).
In the next paragraph, we give a solution for $\xi$ at small $r_s$ and deduce the scaling of  $\Delta E$ from it.

{\bf Analytic solution for small $r_s$.}
For small $r_s$, $\gamma$ is large and  Eq.\ref{eqxi2} can be solved approximately \cite{supp}:
\begin{align}
	\label{ffinal}
	\xi(x)&\approx \frac 1{2\sqrt{\frac{x^2}{x_0^2}+1}}\cos \left(\sqrt{\frac 2{\gamma'}}\arcsinh\left(\frac{x}{x_0}\right)\right)\\
	x_0&=2\exp\left(-\frac{\pi}{2\sqrt 2}\sqrt{\gamma'}-\frac12\right)
	%\T^+\xi(x)&\approx \frac {\pi\gamma \exp(-\sqrt\gamma)}{1+\sqrt\gamma x^2}
\end{align}
for $2\pi\sqrt2(\gamma'-\gamma)=\sqrt\gamma(\pi^2+4)$,
%\begin{align}
%	\label{fsmallx}
%	\xi(x)&\approx \frac12\frac1{\sqrt{ x^2\exp(2\sqrt \gamma)+1}}
%\end{align}
%and for $x>x_0$
%\begin{align}
%	\label{flargex}
%	\xi(x)&\approx \frac12\frac1{x^3\sqrt{ \gamma}\exp(\sqrt \gamma)}
%\end{align}
%provided that $\exp(-\sqrt{\gamma})\ll x_0<\gamma^{-1/4}\ll1$.
%At the fixed point, the energy is given by:
%\begin{align}
%\label{desimple}
%	\de&= -(\T^+\xi,\frac{b^3}{a}).
%\end{align}
leading to the asymptotic behavior of $\Delta E^\F_\SDW$:
\begin{align}
\label{epert}
\Delta E^\F_\SDW(\epsilon)&\lesssim -C\frac{2\pi^2a_V}{r_s} \epsilon^2 \gamma \exp\left(-\frac{\pi}{\sqrt 2}\sqrt {\gamma}\right)%&=\frac{2}{ r_s^2}(\frac 45- {r_s})
\end{align}
with $C=8e^{-3/2-\pi^2/8}$.

Now Eq.\ref{enFG} and Eq.\ref{epert} provide the behavior of $\Delta E(\epsilon,h)$:
\begin{align}
\nonumber
\Delta E&=\Delta E^\F_\SDW+\Delta E^\F_\FG\\
\label{dde}
&=\frac{2\pi^2a_V\gamma \epsilon^2}{r_s}\left(\epsilon\alpha -C\epsfac \right)
%&=\frac{2\pi^2a_V}{r_s}\delta E\\
%\delta E&=\epsilon^3\gamma\alpha -4e^{-3/2}\epsilon^2\gamma\exp\left(-2\sqrt{\gamma}\right)
%\alpha&= \left( \frac23+2h^{2}-2h \right)
%&=\frac{2}{ r_s^2}(\frac 45- {r_s})
\end{align}
\begin{figure}
\begin{center}
\includegraphics[width=0.4\textwidth]{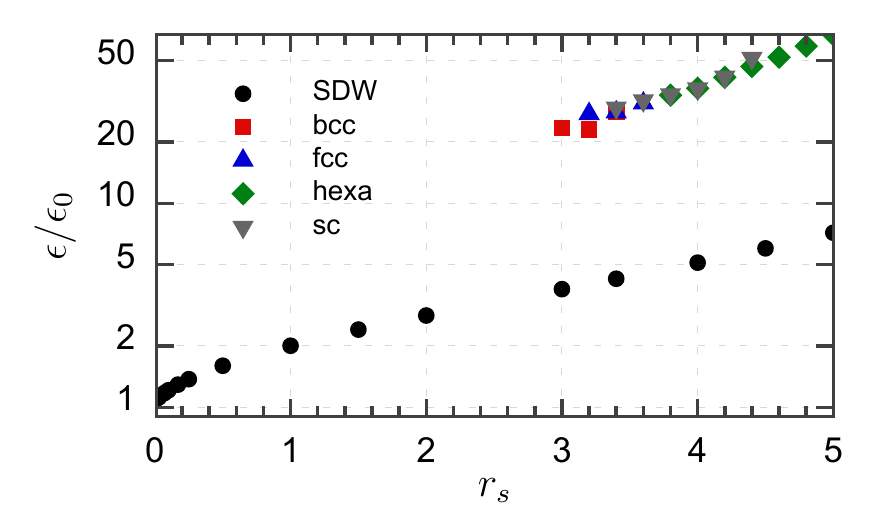}
\caption{Scaled parameter $\epsilon/\epsilon_0$ as a function of the density. $\epsilon_0$  is the value of Eq.\ref{Feps} for $h=1/2$. The black circles stands for the present work, while other symbols stand for HF results\cite{HF3DEG}.% with different symmetries.
\label{ECyl}
}
\end{center}
\end{figure}
The minimum energy is at $\epsilon=\epsilon_0(1+O(\sqrt{r_s}))$:
\begin{align}
\label{DeltaEsmallrs}
\Delta E&\approx -\frac{ \pi a_K\alpha}{r_s^2} \epsilon^3\\
\label{Feps}
\epsilon_0&= \frac{2C}{3\alpha}e^{-\pi^2/4-\pi\sqrt {\gamma_0/2}}\approx \frac{0.0294e^{-7.714/\sqrt {r_s}}}\alpha 
%\epsilon_0&= \frac{2C}{3\alpha}e^{-\pi^2/4}\exp\left(-\left(\frac{3\pi^2}2\right)^{1/3}\frac\pi{\sqrt {r_s}}\right)
\end{align}
where $\gamma_0=a_K/(a_V\pi r_s)$.
%at:
%%\begin{align}
%%\alpha\epsilon\approx \frac 43 \exp(-2\sqrt \gamma)
%%\end{align}
%%that is 
%\begin{align}
%\label{Feps}
%\epsilon \approx \frac{2C}{3\alpha}e^{-\pi^2/4}\exp\left(-\frac{\pi}{\sqrt 2}\sqrt {\gamma_0}\right)
%\end{align}
Eq.\ref{DeltaEsmallrs} shows that at small $r_s$, $\Delta E\, r_s^2/\epsilon^3$ goes to a constant.
Fig.\ref{E} shows the numerical results for the scaled energy at $h=1/2$ ($\blacktriangle$ symbols).
This scaled energy is of order of -0.1 over a wide range of $r_s$.
%This figure shows that $\Delta E$ behaves like $-0.1\epsilon^3r_s^{-2}$ over a large range of $r_s$.
On the other hand, 
while $\epsilon_0$, Eq.\ref {Feps}, varies over decades when $r_s$ decreases, Fig.\ref{ECyl} shows that the ratio $\epsilon/\epsilon_0$ is a slowly varying function. 
% while Fig.\ref{ECyl} shows that  $\epsilon\propto 0.140e^{-7.714/\sqrt{r_s}}$ \cite{supp}. 
The analytical result is supposed to be relevant for large $\gamma$ that is for $ \frac{3}{\sqrt r_s}\gg 1$. This can be verified on the figure: the next corrections in Eqs.\ref{DeltaEsmallrs} and \ref{Feps} behave as $\sqrt{r_s}$. 
%At small $r_s$, $\gamma$ is essentially $\frac{\aK}{\aV \pi r_s}\approx \frac{12.1}{r_s}$.

\begin{figure}
\begin{center}
\includegraphics[width=0.4\textwidth]{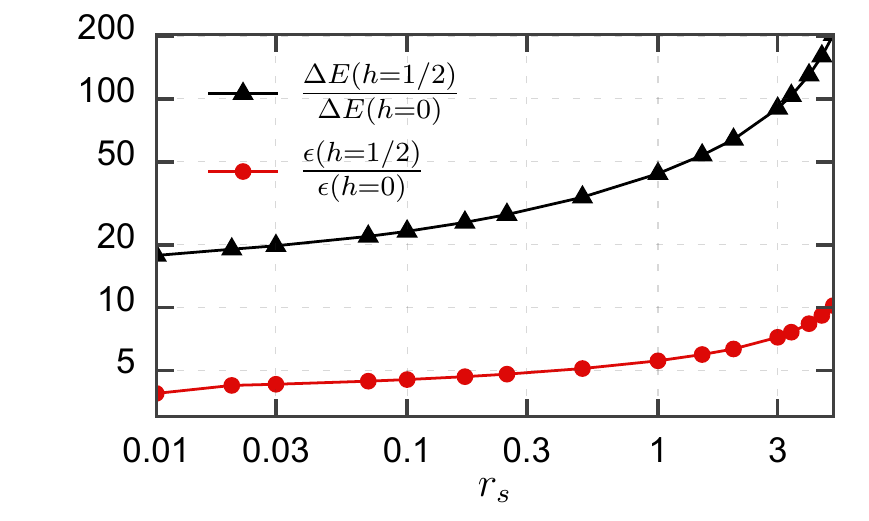}
\caption{Influence of $h$ on the energy gain verus $r_s$:
($\blacktriangle$) energy  and  ($\bullet$) $\epsilon$ ratios.% as a function of $r_s$.  
\label{ERatio}
}
\end{center}
\end{figure}

{\bf Influence of $h$.}
The dependency in $h$ is through the parameter $\alpha$, see Eq.\ref{enFG}.
Fig.\ref{ERatio} shows the effect of $h$ on $\epsilon$ and $\Delta E$ obtained numerically.
For small $r_s$, at $h=1/2$ (thus $\alpha=1/6$), the energy is actually 16 times larger than in the Overhauser model ($h=0$, $\alpha=2/3$).
At larger $r_s$, this ratio can be significantly increased,  e.g. it is about 200 for $r_s=5$;
in this region we expect the energy gain to deeply rely on the precise shape of $\F$.
%For small $r_s$, at $h=1/2$ (thus $\alpha=1/6$), the energy is actually 16 times larger than in the Overhauser model ($h=0$, $\alpha=2/3$) the ratio  is about 200 for $r_s=5$;
%this is a significant effect for numerical simulations.

{\bf Comparison with HF simulations.}
%Fig.\ref{ECyl} shows that $\epsilon$ goes to zero very quickly as $r_s$ decreases.
In previous HF computations of the jellium\cite{HF3DEG,BaguetThesis}  we have considered a discretized Fermi sphere of $64^3$, $96^3$ and $128^3$
values of $\bk$. This corresponds to 32, 48 and 64 equally distributed values of $k$ in the interval $(0,1)$.
For $r_s<5$ evidence for SDW ground states have been found \cite{HF3DEG}. In Figs \ref{E} and \ref{ECyl}, we show the corresponding energy gain per number of SDW (2 for hexa up to 12 for bcc). The larger energy gain of
the HF simulations for $r_s \ge 3$ can be mostly attributed to a smoother and better optimized shape, compared to the simple cylinder
used in the analytical SDW;  other assumptions such as the 1-dimensional approximation decrease the energy further by a factor of $2-3$. For $r_s<3$, the discretization of the Fermi sphere becomes crucial, and even the simulations with $128^3$ k-points used in \cite{HF3DEG} are insufficient to resolve the expected SDW amplitudes leading to the
 standard Fermi gas ground state ($\epsilon=0$ and $b_\bk=0$). Direct numerical simulations of the SDW in this high density region will require a significant increase of k-points by several order of magnitudes.

{\bf Conclusion.}
Considering the ground state of the jellium in the Hartree-Fock approximation, we quantified the energy of the SDW suggested by Overhauser. 
Furthermore, we prove that a modification of the truncated Fermi sphere leads to an energy gain 16 to 200 larger than in the  Overhauser model.

%The effect is very small as the density is large.

Our results readily extends to a polarized model : in Eq.\ref{potSDW} we have to take into account the direct potential which appears with a factor $1/Q^2$ and thus is negligible at small $r_s$ (of order $\epsilon^4$).

In order to obtain the energy of jellium, the results of Fig.\ref{ECyl} must be multiplied by the number of SDW.
For simple periodic models considered in previous works, this factor varies from 2 (hexa) up to 12 (bcc).
At very small $r_s$, the perturbation of the SDW is localized  in tiny regions which do not interact,
thus, one may suppose that we can have many of them distributed around the Fermi sphere giving rise to a quasi-periodic behavior
of the density.


\begin{thebibliography}{99}
\bibitem{bpositive} One easily checks afterwards that the optimal energy is obtained for $b_\bk$'s
with a constant arbitrary phase.
\end{thebibliography}
\end{document}